\documentclass[a4paper,12pt]{article}
\usepackage{graphicx}
\usepackage[english]{babel}
\usepackage{amssymb,amsbsy,amsmath,eucal}
\usepackage[compress]{cite}
\newcommand{\lab}[1]{\label{#1}}
\newcommand{\re}[1]{(\ref{#1})}
\newcommand{\nn}{\nonumber}
\newcommand{\B}[1]{\boldsymbol{#1}}

\newcommand{\sO}{\mathsf{O}}
\newcommand{\dO}{\dot{\sO}}
\newcommand{\BOm}{\boldsymbol{\Omega}}
\newcommand{\Ec}{\B{\cal E}}
\newcommand{\Pc}{\B{\cal P}}

\newcommand{\D}[2]{{\rm d}^{#1}{#2}\,}
\begin{document}

\title{Free rotation of conducting and dielectric spheres in a uniform electrostatic field}

\author{A. Duviryak}
\date{Institute for Condensed Matter Physics of NAS of Ukraine,\\
1 Svientsitskii Street, Lviv, UA-79011, Ukraine \\
              Tel.: +380 322 701496, \
              Fax: +380 322 761158\\
              {duviryak@icmp.lviv.ua}           
}

\maketitle

\begin{abstract}
Rotation of conducting and dielectric spherical particles levitating
in the uniform electrostatic field is considered. A dipole moment of
the spherical particle induced by the external uniform electrostatic
field is inclined to the field if the particle rotates. This causes
the torque braking the rotation. Vectors of dipole moment and torque
depend both on an angular velocity of the particle and its electric
properties. Equations of rotary motion of the particle levitating in
the external field are integrable in quadratures. Few examples of
the conducting and dielectric particles are solved explicitly.
\end{abstract}


\section{Introduction}
\renewcommand{\theequation}{1.\arabic{equation}}
\setcounter{equation}{0}

A uniform electrostatic field induces in a motionless spherical
conducting or dielectric particle a dipole moment which is parallel
or antiparallel to the field. If the particle is immersed in a
conducting fluid, the interaction of the field with the induced
dipole moment may cause spontaneous rotation of the particle, known
as the Quincke effect \cite{M-T69,Jon84}. On the contrary, if the
particle is already rotating in a vacuum or a non-conducting medium,
the induced dipole moment becomes inclined to the field \cite{Jon95}
and gives rise to the torque braking the particle rotation. The
magnitude and the direction of the induced dipole moment depend on
the angular velocity of the particle as well as on its electric
properties. The same is concerned with the induced torque braking
the rotation. Nowadays, nanoparticles in optical traps can be spun
up to GHz, and the rotary frequency is in rapid progress
\cite{RDHD18,AXBJGL20,JYRLYZ21}. The question arises naturally: may
the aforementioned effects become essential in future experiments\,?

Here the rotation of neutral spherical particles levitating in the
uniform electrostatic field under the action of the induced braking
torque is analyzed. Both the conductive and dielectric particles are
considered in Sections 2 and 3, respectively. The study is
two-stage. On the first stage the electrostatic potential of a
rotating particle and a corresponding induced dipole moment is
calculated. This task requires a generalization in Subsections
2.1--2.3 of results presented by T.B. Jones \cite{Jon84,Jon95} to
the case where the axis of particle rotation is oriented arbitrarily
with respect to the external field. Besides, the dispersion of
dielectric permittivity must be taken into account in Subsection 3.1
when considering dielectric particles. On the second stage
expressions for the torque of a dipole interaction with the external
field is used to formulate and analyze the equations of rotary
motion. They are shown integrable in quadratures for both conductive
and dielectric particles; see Subsections 2.4 and 3.2, respectively.
Few basic examples has been solved analytically. They are the Ohm
conductive particle (Section 2, Subsection 2.4), the Debye model of
polar dielectric (Subsection 3.3), and the Lorentz model of
non-polar  dielectric (Subsection 3.4). Models can be complicated
and combined. The particles with hybrid conductor-dielectric
properties are considered in Section 4. All the models are supported
by corresponding numerical examples in Subsections 2.5, 3.3.1, 3.4.1
and 4.1.

Conclusion and a possible application of presented results in the
particle trap physics is discussed in Section 5.

All analytical calculations are presented in the CGS system. Some
input and output data in the numerical examples are given in
commonly used non-CGS units.


\section{Spherical conductive particle rotating in the uniform electrostatic field}
\renewcommand{\theequation}{2.\arabic{equation}}
\setcounter{equation}{0}

\subsection{Electrostatics of a rotating conductor}

Let us consider an electrically neutral rigid particle rotating in
vacuum. The particle consists of a uniform conducting medium of the
volume charge density $\rho$ and the current density $\B j$ in the
laboratory reference frame. We are interested of an electric
response of the particle on an external uniform electrostatic field,
i.e., the charge redistribution and the local electric field inside
and around the particle. We suppose that relativistic effects are
negligible. Thus one can apply the set of quasi-static electric
field equations \cite[Table\,I]{M-T69} to our case of the rotating
conductor.

We describe the rotary motion of the particle by means of the
time-dependent matrix $\sO(t)\in\,$SO(3): $\B r'=\sO(t)\B r(t)$,
where $\B r(t)$ is a radius-vector of any material point of particle
in the laboratory reference frame, and $\B r'$ is a corresponding
constant vector in the proper reference frame (where the particle is
motionless). We will assume that in the proper reference frame the
electric field $\B E'=\sO\B E$  and the current density $\B j'$ are
related by the local Ohm law:
%
\begin{equation}
\B j'=\varkappa\B E', \lab{2.1}
\end{equation}
where the conductivity $\varkappa$ is a positive constant. Then in
the laboratory reference frame the current density includes the
conduction part and the convection part \cite{M-T69}:
%
\begin{equation}
\B j=\varkappa\B E+\rho\,\B v \lab{2.2}
\end{equation}
where
%
\begin{equation}
\B v=-\sO^{\rm T}\dO\B r=\BOm\times\B r \lab{2.3}
\end{equation}
is the velocity of the point $\B r$, $\dO\equiv\D{}{{\sO}}/\D{}{t}$,
the angular velocity vector $\BOm$ is dual to the skew-symmetric
matrix $\sO^{\rm T}\dO$, and ``~$\times$~'' denotes the cross
product of vectors.

We suppose that despite of rotation the particle is in a steady
state, at least adiabatically (this term will be clarified farther).
Thus
%
\begin{equation}
\partial\rho/\partial t=0\qquad\Longrightarrow\qquad \nabla\cdot\B j=0 ,
\lab{2.4}
\end{equation}
and the equations for the electrostatic potential $\varphi$ hold:
%
\begin{subequations}\lab{2.5}
\begin{eqnarray}
\B E&=&-\nabla\varphi,
\lab{2.5a}\\
\nabla\cdot\B E&=&-\Delta\varphi=4\pi\rho. \lab{2.5b}
\end{eqnarray}
\end{subequations}
The equations \re{2.2}--\re{2.5} yield the equation for the charge
density in the bulk of particle:
%
\begin{equation}
\BOm\cdot(\B r\times\nabla)\rho+4\pi\varkappa\rho=0. \lab{2.6}
\end{equation}

The bulk equations \re{2.4}--\re{2.6} are complemented with the
boundary conditions:
%
\begin{eqnarray}
E_{n}^+-E_{n}^-&=&4\pi\sigma,
\lab{2.7}\\
\varphi^+-\varphi^-&=&0, \lab{2.8}\\
\nabla\cdot\B i - j^-_n&=&0; \lab{2.9}
\end{eqnarray}
here $\sigma$ and $\B i$ are the charge and the current surface
densities, the superscript ``$-$'' denotes the surface approach from
the internal medium side, the superscript ``+'' denotes the approach
from the external vacuum side, and the unit normal $\B n$ is
directed outside the particle surface so that $j^+_n\equiv\B
n\cdot\B j^+=0$.

\subsection{Charge balance equations}

We consider a spherical particle of the radius $R$ and the
conductivity $\varkappa$ rotating with the constant angular velocity
$\BOm$ in the external uniform electrostatic field. We denote this
field by the constant vector $\Ec$ to distinguish from the
aforementioned local electric field $\B E$. Let $\BOm$ be directed
along the orth $\B e_z$ of the Cartesian coordinate system O$xyz$ in
the laboratory reference frame, i.e., $\BOm=\Omega\,\B e_z$. This
can be done by the following choice of the matrix
$\sO(t)\in\,$SO(3):
%
\begin{equation}\lab{2.10}
\sO(t)=\left[\begin{array}{ccc}
~\cos\Omega\,t~~ & ~~\sin\Omega\,t~~ & 0 \\
-\sin\Omega\,t~~  &  ~~\cos\Omega\,t~~ & 0 \\
0  & 0 & 1
\end{array}\right].
\end{equation}
Then we use the spherical coordinate system built from the center of
the particle,
%
\begin{eqnarray}
&&x=r\sin\theta\cos\alpha,\quad y=r\sin\theta\sin\alpha,\quad z=r\cos\theta,\nn\\
&&0\le r<\infty,\quad\qquad 0\le \theta\le\pi,\quad\qquad 0\le
\alpha<2\pi \lab{2.11}
\end{eqnarray}
and endowed by the right orth triplet $\B e_r$, $\B e_\theta$, $\B
e_\alpha$ at every point $\B r$; Fig. \ref{fig1}.
%
\begin{figure}[ht]
\begin{center}
\includegraphics[scale=0.4]{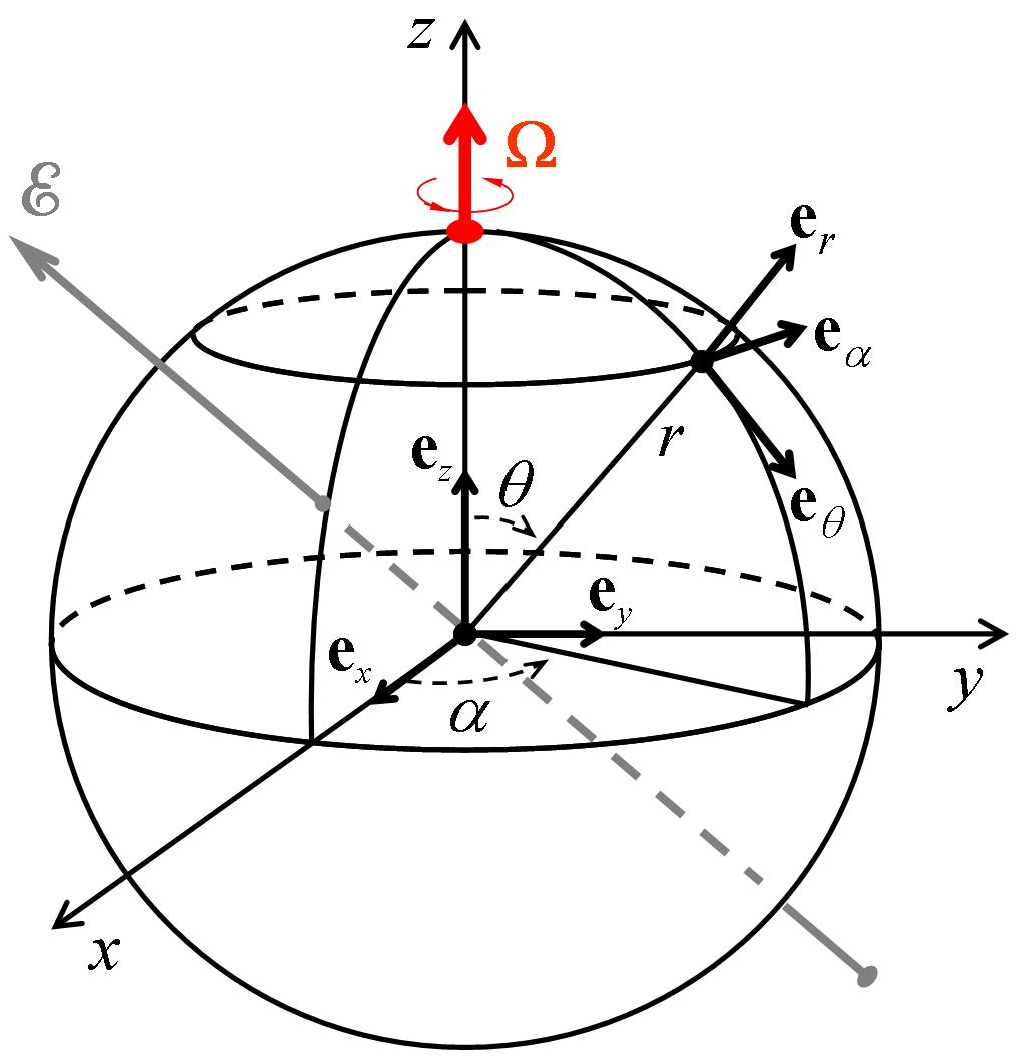}
\caption{The choice of the spherical coordinate system convenient to
derive the polarization of a rotating particle in the uniform
electric field $\Ec$. }\lab{fig1}
\end{center}
\end{figure}
%

Then the equation \re{2.6} takes the form
%
\begin{equation}
\Omega\partial\rho/\partial\alpha+4\pi\varkappa\rho=0,
\end{equation}
and the only continuous solution is $\rho=0$, as in the static case
$\Omega=0$. Nevertheless, the charge can accrue in the surface layer
of some small thickness $h$ yielding the surface density
$\sigma=h\rho|_{r=R}$.

In contrast, the current can flow along both the bulk and surface,
so that
%
\begin{eqnarray}
\B j&=&j_r\B e_r+j_\theta\B e_\theta+j_\alpha\B e_\alpha=\varkappa\B
E,
\lab{2.13}\\
\B i&\equiv&h\B j|_{r=R}=i_\theta\B e_\theta+i_\alpha\B
e_\alpha=[\kappa\B E_{\rm T}+\sigma\BOm\times\B r]_{r=R}, \lab{2.14}
\end{eqnarray}
where $\B i$ is a surface current density, $\B E_{\rm T}\equiv\B
E-E_n\B n=E_\theta\B e_\theta+E_\alpha\B e_\alpha$ is the tangential
projection of $\B E$ to the particle surface, and $\kappa\equiv
h\varkappa$; but one can assume in general $\kappa\ne h\varkappa$ if
electric properties of the surface layer differ from the bulk ones.
For example, one can put $\varkappa R\lesssim\kappa$ if a
poor-conducting particle is covered with the well-conducting layer,
or $\varkappa=0$ if the core is dielectric. This example will be
considered in the Section 4, Subsection 4.1.2.

The charge conservation law \re{2.4} is fulfilled  in the bulk by
the equations \re{2.2}, \re{2.5b} with $\rho=0$ in r.-h.s. For the
surface layer one should write down the boundary condition \re{2.9}
in spherical coordinates. It is illustrative to treat \re{2.9} as an
application of the conservation law \re{2.4} in the integral form,
where the integration runs over a surface of a small volume segment
of the layer; Fig. \ref{fig2}:
%
\begin{eqnarray}
\oint\D{}{\B s}\cdot\B j=-j_rab +[\tilde
bj_{\theta+\delta\theta}-bj_{\theta}]h
+\left[j_{\alpha+\delta\alpha}-j_\alpha\right]ha&=&0, \lab{2.15}
\end{eqnarray}
where~~$a=R\delta\theta,~~b=R\sin\theta\,\delta\alpha,~~\tilde
b=R\sin(\theta+\delta\theta)\,\delta\alpha$, ~and $h$ is a thickness of the layer.\\
One arrives at the equation:
%
\begin{equation}
\frac{\partial (i_\theta\sin\theta)}{\partial\theta} +
\frac{\partial i_\alpha}{\partial\alpha} - Rj_r^-\sin\theta=0
\lab{2.16}
\end{equation}
where $j_r^-$ is the radial component of the volume current density
just below the surface layer.
%
\begin{figure}[ht]
\begin{center}
\includegraphics[scale=0.6]{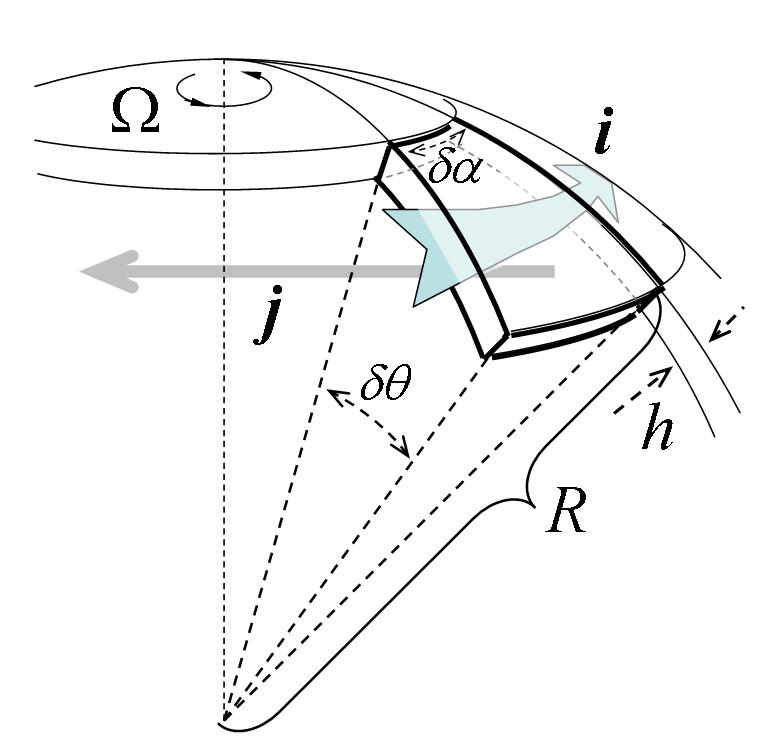}
\caption{Current balance in a surface layer of the rotating
particle; visualization of the equations \re{2.13}--\re{2.16}
}\lab{fig2}
\end{center}
\end{figure}
%

\subsection{Electrostatic potential}

Up to this point a solution of the electrostatic equations \re{2.5}
and boundary conditions \re{2.7}, \re{2.8} including the surface
charge density $\sigma$ remained unknown. From now on we imply the
surface layer negligibly thin, $h\to0$, and use the following ansatz
for the electrostatic potential:
%
\begin{subequations}\lab{2.17}
\begin{align}
\varphi&=-\Ec\cdot\B r + \frac{R^3}{r^3}\,\Pc\cdot\B r,   &r>R,
\lab{2.17a}\\
\varphi&=-(\Ec-\Pc)\cdot\B r,   &r<R, \lab{2.17b}
\end{align}
\end{subequations}
where $\Pc$ is an unknown constant vector to be found. This
potential satisfies the boundary condition \re{2.8}. The 1st term in
r.-h.s of \re{2.17a} is the potential of the external field $\Ec$,
and the 2nd one is the potential of the dipole moment $\B d=R^3\Pc$.
Was $\Pc=\Ec$, one arrives at the static case $\Omega=0$.

It follows from \re{2.17} the field on the surface:
%
\begin{eqnarray*}
\B E^+&\equiv&\B E|_{r\to R+0}=\Ec - \Pc + 3{\cal P}_r\B e_r,
\\
\B E^-&\equiv&\B E|_{r\to R-0}=\Ec - \Pc,
\end{eqnarray*}
then the boundary condition \re{2.7} yields the surface charge
density:
%
\begin{equation}
\sigma=\frac1{4\pi}(E_r^+-E_r^-)=\frac3{4\pi}{\cal P}_r. \lab{2.18}
\end{equation}

Relations \re{2.11} between Cartesian and spherical coordinates
yield relations between corresponding vector components, for one:
%
\begin{eqnarray}
\Ec=\mathcal{E}_x\B e_x+\mathcal{E}_y\B e_y+\mathcal{E}_z\B e_z\equiv\mathcal{E}_r\B e_r+\mathcal{E}_\theta\B e_\theta+\mathcal{E}_\alpha\B e_\alpha\nn\\
=(\mathcal{E}_x\sin\theta\cos\alpha+\mathcal{E}_y\sin\theta\sin\alpha+\mathcal{E}_z\cos\theta)\B e_r\nn\\
{}+(\mathcal{E}_x\cos\theta\cos\alpha+\mathcal{E}_y\cos\theta\sin\alpha-\mathcal{E}_z\sin\theta)\B e_\theta\nn\\
{}+(-\mathcal{E}_x\sin\alpha+\mathcal{E}_y\cos\alpha)\B e_\alpha
 \lab{2.19}
\end{eqnarray}
etc. Gathering all expressions involved in \re{2.13}--\re{2.14} and
then requiring the equality \re{2.16} at every point of the surface,
one arrives at the set of equations for components of the vector
$\Pc$:
%
\begin{eqnarray}
\left(\varkappa+2\kappa/R\right)(\mathcal{E}_x&-&{\cal P}_x)=\frac{3\Omega}{4\pi}{\cal P}_y,\nn\\
\left(\varkappa+2\kappa/R\right)(\mathcal{E}_y&-&{\cal P}_y)=-\frac{3\Omega}{4\pi}{\cal P}_x,\nn\\
\mathcal{E}_z&-&{\cal P}_z=0. \lab{2.20}
\end{eqnarray}
In view of the spherical symmetry of the problem one can put
$\mathcal{E}_y=0$ without loss of generality. Then the solution of
the set \re{2.20} is
%
\begin{equation}
{\cal P}_x=\frac{\mathcal{E}_x}{1+\tau_0^2\Omega^2}, \qquad {\cal
P}_y=\frac{\tau_0\Omega\mathcal{E}_x}{1+\tau_0^2\Omega^2}, \qquad
{\cal P}_z=\mathcal{E}_z, \lab{2.21}
\end{equation}
where
%
\begin{equation}
\tau_0=\frac{3}{4\pi(\varkappa+2\kappa/R)} \lab{2.22}
\end{equation}
is the characteristic time scale, during which the charge
distribution follows the change of the particle orientation with
respect to the external field. In particular,
$\Pc\mathop{\longrightarrow}\limits_{\Omega\to0}\Ec$, as it is
expected in the static case, so the field inside the conducting
sphere disappears, $\B E=0$. On the other hand, ${\cal
P}_x\mathop{\longrightarrow}\limits_{\Omega\to\infty}0$, ${\cal
P}_y\mathop{\longrightarrow}\limits_{\Omega\to\infty}0$, i.e., the
charge distribution cannot follow a fast rotation and smears over
the surface.

It is convenient to represent the solution \re{2.21} in
coordinateless form. For this purpose we recall the definition of
the Cartesian orth triplet according to Fig. \ref{fig1}:
%
\begin{eqnarray}
\BOm\|\B e_z\qquad\Longrightarrow\qquad\B e_z&=&\BOm/\Omega,\qquad \mbox{where}\quad \Omega=|\BOm|; \nn\\
\Ec\in{\rm O}xz\qquad\Longrightarrow\qquad\B e_x&=&\Ec_\bot/\mathcal{E}_\bot, \quad \mbox{where}\quad \Ec_\bot=\Ec-(\Ec\cdot\BOm)\BOm/\Omega^2\nn\\
\B e_y&=&\B e_z\times\B e_x. \lab{2.23}
\end{eqnarray}
Then, using \re{2.21} and \re{2.23} we have:
%
\begin{eqnarray}
\Pc&=&{\cal P}_x\B e_x+{\cal P}_y\B e_y+{\cal P}_z\B
e_z=\frac1{1+\tau_0^2\Omega^2}\,\{\Ec+\tau_0\BOm\times\Ec+\tau_0^2(\BOm\cdot\Ec)\BOm\}.
\lab{2.24}
\end{eqnarray}

Equations \re{2.17}, \re{2.24} represent 3D generalization of the 2D
electrostatic potential known for the particular case $\Ec\bot\BOm$
\cite{Jon84,Jon95} and used for a description of the Quincke effect.

\subsection{Rotary dynamics}

It follows from the previous section that the spherical conductor
rotating in the static uniform electric field acquires the dipole
moment $\B{d}=R^3\B{\cal P}$ which interaction with the external
field $\Ec$ changes the angular momentum of the
particle as follows
%
\begin{equation}
I\frac{\D{}{\BOm}}{\D{}{t}}=\B
M\equiv\B{d}\times\Ec=R^3\Pc\times\Ec, \lab{2.25}
\end{equation}
where $I$ is the inertia moment of the spherical particle. The
equation is somewhat inconsistent since the solution \re{2.17},
\re{2.24} was built for a steady state configuration, provided
$\Ec$, $\BOm$ and thus $\Pc$ are constant vectors. Actually, we
adopt the adiabatic approximation, i.e., we assume that the charge
and current distributions follow immediately or very quickly the
change of the angular velocity $\BOm(t)$, i.e., the vector
$\Pc(t)=\Pc[\BOm(t)]$.

Inserting this expression into the equation \re{2.25} reduces the
latter to the form of non-linear Euler equation:
%
\begin{equation}
I\frac{\D{}{\BOm}}{\D{}{t}}=\frac{\tau_0R^3}{1+\tau_0^2\Omega^2}\Ec\times\{\Ec\times\BOm-\tau_0(\BOm\cdot\Ec)\BOm\}.
\lab{2.26}
\end{equation}
It is convenient to split this vector equation by means of new
Cartesian coordinate system spanned on the orth triplet $\B e_1$,
$\B e_2$, $\B e_3$ such that $\Ec=\mathcal{E}\B e_3$; Fig.
\ref{fig3}.
%
\begin{figure}[ht]
\begin{center}
\includegraphics[scale=0.4]{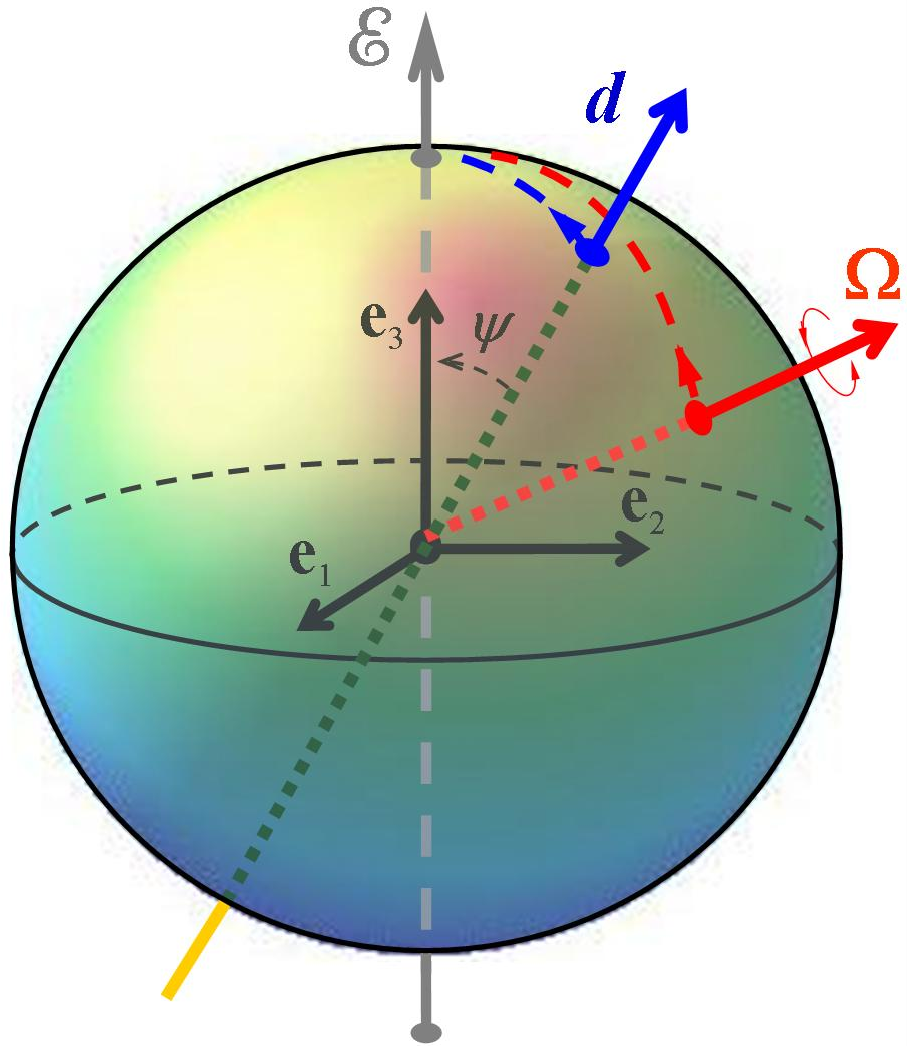}
\caption{The choice of the Cartesian coordinate system convenient to
describe a rotary dynamics of the particle in the uniform electric
field $\Ec$. }\lab{fig3}
\end{center}
\end{figure}
%

Introducing dimensionless variables:
%
\begin{equation}
\B\omega=\tau_0\BOm, \qquad \tau=t/T_0, \lab{2.27}
\end{equation}
where the time scale parameter $\tau_0$ is defined by \re{2.22}, and
%
\begin{equation}
T_0=I/(\tau_0R^3\mathcal{E}^2), \lab{2.28}
\end{equation}
reduces \re{2.26} to the dimensionless component set:
%
\begin{eqnarray}
\frac{\D{}{\omega_1}}{\D{}{\tau}}&=&-\frac{\omega_1-\omega_3\omega_2}{1+\omega^2},
\lab{2.29}\\
\frac{\D{}{\omega_2}}{\D{}{\tau}}&=&-\frac{\omega_2+\omega_3\omega_1}{1+\omega^2},
\lab{2.30}\\
\frac{\D{}{\omega_3}}{\D{}{\tau}}&=&0. \lab{2.31}
\end{eqnarray}
Thus $\omega_3=\,$const, and the vector
$\B\omega_\bot\equiv\{\omega_1,\omega_2,0\}$ is to be found.

Adding \re{2.29} multiplied by $\omega_1$ with \re{2.30}  multiplied
by $\omega_2$ yields the equation for
$\omega_\bot^2\equiv\B\omega_\bot\cdot\B\omega_\bot=\omega_1^2+\omega_2^2$:
%
\begin{equation*}
\frac{\D{}{\omega_\bot^2}}{\D{}{\tau}}=-\frac{2\omega_\bot^2}{1+\omega_3^2+\omega_\bot^2}
\end{equation*}
which integration results in the formula:
%
\begin{equation}
\ln\nu+\nu=\ln\nu_0+\nu_0-2\tau/\xi, \lab{2.32}
\end{equation}
where
%
\begin{equation*}
\nu\equiv\omega_\bot^2/\xi, \qquad \xi=1+\omega_3^2,\qquad
\nu_0=\nu|_{\tau=0}.
\end{equation*}
This implicit dependency of $\omega_\bot^2$ on $\tau$ can be
presented explicitly by means of the Lambert W-function
\cite{Cor96}:
%
\begin{equation}
\nu=W_0\left(\nu_0{\rm e}^{\nu_0-2\tau/\xi}\right).\lab{2.33}
\end{equation}
Now introducing the subsidiary evolution parameter
%
\begin{equation}
\lambda=\ln\sqrt{\nu_0/\nu}, \qquad
\lambda\sim\frac\tau{1+\omega_3^2}\quad\mbox{at}\quad\tau\to\infty
\lab{2.34}
\end{equation}
reduces the set \re{2.29}-\re{2.30} to the linear equations with
constant coefficients:
%
\begin{eqnarray*}
\D{}{\omega_1}/\D{}{\lambda}&=&-\omega_1-\omega_3\omega_2,
\\
\D{}{\omega_2}/\D{}{\lambda}&=&-\omega_2+\omega_3\omega_1,
\end{eqnarray*}
which solution reads:
%
\begin{equation}
\omega_1=\omega_{10}{\rm e}^{-\lambda}\cos\phi, \quad
\omega_2=\omega_{10}{\rm e}^{-\lambda}\sin\phi, \quad
\phi=-\omega_3\lambda. \lab{2.35}
\end{equation}
Here, using the spherical symmetry, we put
$\omega_{20}\equiv\omega_2(0)=0$ so that
$\omega_{\bot0}=|\omega_{10}|$.

The vector $\Pc$ and thus the dipole moment $\B{d}$ are inclined to
the direction $\B e_3$ of the vector $\Ec$ by the angle $\psi$ (see
Figure \ref{fig3}):
%
\begin{equation}
\tan\psi=\omega_\bot/\sqrt{1+\omega_3^2} \lab{2.36}
\end{equation}
which is maximal if $\omega_3=0$.

Coming back to the dimensional variables, let us present the
solution \re{2.35} asymptotically, at $t\to\infty$:
%
\begin{equation}
\Omega_\bot\sim{\rm e}^{-t/T}, \qquad \phi\sim-\tilde\Omega_3t,
\lab{2.37}
\end{equation}
where
%
\begin{equation}
T=T_0(1+\tau_0^2\Omega_3^2), \qquad \tilde\Omega_3=\Omega_3\tau_0/T.
\lab{2.38}
\end{equation}
It follows from \re{2.37} that the transversal to $\Ec$ component
$\BOm_\bot$ of the angular velocity $\BOm$ decreases in its
magnitude exponentially with the braking time $T$ and precesses
clockwise with the angular velocity $\tilde\Omega_3$. The
inclination angle $\psi\to0$ by the same exponential law.

\subsection{Numerical examples}

If the spherical particle is homogeneous the mass and the inertia
moment are:
%
\begin{equation*}
m=\frac43\pi\mu R^3, \qquad I= \frac25mR^2=\frac8{15}\pi\mu R^5
\end{equation*}
where $\mu$ is a mass density. Besides, the surface conductivity is
negligible compared to the effect of the bulk one:
$\kappa/(\varkappa R)=h/R\to0$ . Then two time scale parameters
\re{2.22} and \re{2.28} present in the problem specify as follows:
%
\begin{eqnarray*}
\tau_0&=&\frac3{4\pi\varkappa}, \qquad
T_0=\frac{32\pi^2}{45}\,\frac{\mu\varkappa R^2}{\mathcal{E}^2}.
\end{eqnarray*}
The first one does not depend on the size of particle. It determines
the relaxation time in which the charge distribution of the particle
follows its rotary motion. The second scale parameter, actually, is
$T=T_0(1+\tau_0^2\Omega_3^2)$; it determines the characteristic time
in which $\Omega_\bot\to0$, as it follows from \re{2.37}. The
approach of the vector $\BOm_\bot\to0$ and thus $\Pc\to\Ec$ is
accompained  by a precession with the angular velocity
$\tilde\Omega_3=\Omega_3\tau_0/T$, as it follows from the solution
\re{2.35}, \re{2.34} or \re{2.37}, \re{2.38}.

Let us recall that this solution corresponds to the adiabatic
approximation which in present case is valid provided $\tau_0\ll T$.

Here we consider two examples of metallic particles: the golden
particle (the perfect conductor) and the nichrome particle (higher
resistance conductor). In these and subsequent examples the size and
the maximal angular velocity will be put as in the experiment
\cite{RDHD18}: $R=50\,$nm, $\Omega_{\rm max}=2\pi\,$GHz. We put the
external field $\mathcal{E}=33.4\,$statV/cm (i.e., $10^6\,$V/m in SI
units) which is characteristic for devices like the Van der Graaf
electrostatic generator e.t.c. This value provides an essential
induced dipole moment of conductive particles ${d} \approx
R^3\mathcal{E} \approx 4400\,$Debyes =
4.4$\cdot10^{-15}\,$statC$\cdot$cm, the same value as the permanent
dipole moment of the polar cellulose nanocrystals \cite{F-PBL14}.

\subsubsection{Perfect conductor: golden nanoparticle}
$\mu=19.31\,$g/cm$^3$, $\varkappa=36.9\cdot10^{16}\,$s$^{-1}$ (i.e.,
4.1$\cdot10^{7}\,$S/m in SI). Then $\tau_0=6.37\cdot10^{-19}\,$s,
and $\omega\le\omega_{\rm max}=\tau_0\Omega_{\rm
max}=4\cdot10^{-9}$. Thus the braking time $T\approx
T_0=1.14\cdot10^6\,$s$\ \approx13$ days is of order of a storage
time in Penning trap \cite{Haf03}. Since $T\gg\tau_0$, the adiabatic
approximation is perfect. The initial inclination angle is not more
than $\psi\lesssim4\cdot10^{-9}$, so the effect is negligible.

\subsubsection{High resistance conductor: nichrome nanoparticle}
$\mu=8.5\,$g/cm$^3$, $\varkappa=9\cdot10^{15}\,$s$^{-1}$ (i.e.,
$10^{6}\,$S/m in SI). Then $\tau_0=2.66\cdot10^{-17}\,$s, and
$\omega\le\tau_0\Omega_{\rm max}=1,7\cdot10^{-7}$. Thus the
inclination is negligible, $\psi\lesssim1.7\cdot10^{-7}\,$, and
$T\approx T_0=1.2\cdot10^4\,$s$\ \approx3.34$ hours. Again,
$T\gg\tau_0$, and the adiabatic approximation is perfect.


\section{Spherical dielectric particle rotating in the uniform electrostatic field}
\renewcommand{\theequation}{3.\arabic{equation}}
\setcounter{equation}{0}

\subsection{Electrostatics of a rotating dielectric}

We assume that in the proper reference frame the material equation
relating the electric field $\B E'$ and the electric induction $\B
D'$ is linear, isotropic but time-nonlocal:
%
\begin{equation}
\B D'(t)=\int\limits_{-\infty}^{t}\D{}{t'}\epsilon(t-t')\B E'(t').
\lab{3.1}
\end{equation}
The transition to the laboratory reference frame, $\B E'(t)=\sO(t)\B
E$, where the electric field $\B E$ is again assumed static, yields
thus the static but anisotropic relation:
%
\begin{equation}
\B D=\hat\varepsilon\B E, \lab{3.2}
\end{equation}
where the tensor of dielectric permittivity
%
\begin{equation}
\hat\varepsilon=\int\limits_{0}^{\infty}\D{}{t}\epsilon(t){\sO}^{\rm
T}(t), \lab{3.3}
\end{equation}
is determined via the kernel $\epsilon(t)$ and the rotation matrix
\re{2.10} transposed $\sO^{\rm T}(t)$. Explicitly,
%
\begin{equation}\lab{3.4}
\hat\varepsilon=\left[\begin{array}{ccc}
\varepsilon_{\rm r} & -\varepsilon_{\rm i} & 0 \\
\varepsilon_{\rm i}  &  \varepsilon_{\rm r} & 0 \\
0  & 0 & \varepsilon_{\rm s}
\end{array}\right],
\end{equation}
where
%
\begin{eqnarray}
\varepsilon_{\rm r}\equiv\mathrm{Re}\,\varepsilon(\Omega),\qquad
\varepsilon_{\rm i}\equiv\mathrm{Im}\,\varepsilon(\Omega),\qquad
\varepsilon_{\rm s}\equiv\varepsilon(0), \qquad
\varepsilon(\Omega)\equiv\int\nolimits_{0}^{\infty}\D{}{t}\epsilon(t)\mathrm{e}^{\mathrm{i}\Omega
t}, \lab{3.5}
\end{eqnarray}
$\mathrm{Re}\,\varepsilon(\Omega)$ and
$\mathrm{Im}\,\varepsilon(\Omega)$ are the real and imaginary parts
of the dielectric function $\varepsilon(\Omega)$ to be the Fourier
transform of the kernel $\epsilon(t)$, and $\varepsilon_{\rm s}$ is
the static dielectric constant.

We will use the same ansatz \re{2.17} for the electrostatic
potential with the polarization vector $\Pc$ to be found, but the
boundary condition \re{2.7} will be reformulated in terms of the
electric induction field:
%
\begin{equation}
E_r^+=D_r^-. \lab{3.6}
\end{equation}
Using the expansions like \re{2.19} one can recast the condition
\re{3.6} into the following equation for the polarization vector
$\Pc$:
%
\begin{equation*}
\Ec+2\Pc=\hat\varepsilon(\Ec-\Pc)
\end{equation*}
which can be solved in the symbolic form:
%
\begin{equation}
\Pc=[\hat\varepsilon+2]^{-1}(\hat\varepsilon-1)\Ec
\end{equation}
revealing the matrix analogue of the Clausius-Mossotti relation.

Using the spherical symmetry and putting (as before)
$\mathcal{E}_y=0$, one arrives at the explicit expressions:
%
\begin{eqnarray}
{\cal P}_x=\frac{(\varepsilon_{\rm r}+2)(\varepsilon_{\rm
r}-1)+\varepsilon_{\rm i}^2}{(\varepsilon_{\rm
r}+2)^2+\varepsilon_{\rm i}^2}\,\mathcal{E}_x,\qquad {\cal
P}_y=\frac{3\varepsilon_{\rm i}}{(\varepsilon_{\rm
r}+2)^2+\varepsilon_{\rm i}^2}\,\mathcal{E}_x,\qquad {\cal
P}_z=\frac{\varepsilon_{\rm s}-1}{\varepsilon_{\rm
s}+2}\,\mathcal{E}_z. \lab{3.8}
\end{eqnarray}

\subsection{Rotary dynamics}

Combining the formulae \re{3.8} with the Cartesian orth triplet
\re{2.23} and inserting the dipole moment $\B{d}=R^3\Pc$ into the
r.-h.s. of \re{2.25} one obtains the Euler equation
%
\begin{eqnarray}
I\frac{\D{}{\BOm}}{\D{}{t}}=\frac{R^3}{\Omega^2}\Ec\times\left\{
\frac{3\varepsilon_{\rm i}\Omega\,\Ec\times\BOm}{(\varepsilon_{\rm
r}+2)^2+\varepsilon_{\rm i}^2} - \left[\frac{\varepsilon_{\rm
s}-1}{\varepsilon_{\rm s}+2}- \frac{(\varepsilon_{\rm
r}+2)(\varepsilon_{\rm r}-1)+\varepsilon_{\rm
i}^2}{(\varepsilon_{\rm r}+2)^2+\varepsilon_{\rm i}^2}\right]
(\BOm\cdot\Ec)\BOm\right\}. \lab{3.9}
\end{eqnarray}
Choosing the dimensionless variables \re{2.27},  \re{2.28}, where
$\tau_0$ is some characteristic for a dielectric relaxation time
(instead of \re{2.22} for a conductor), and splitting the Euler
equation by components, we have again $\omega_3=\,$const, and arrive
at the set of nonlinear equations for $\omega_1$ and $\omega_2$:
%
\begin{eqnarray}\lab{3.10}
\frac{\D{}{\omega_1}}{\D{}{\tau}}&=&-\frac{f(\omega^2)}{\omega^2}\omega_1 + \frac{g(0)-g(\omega^2)}{\omega^2}\omega_3\omega_2, \nn\\
\frac{\D{}{\omega_2}}{\D{}{\tau}}&=&-\frac{f(\omega^2)}{\omega^2}\omega_2
- \frac{g(0)-g(\omega^2)}{\omega^2}\omega_3\omega_1,
\end{eqnarray}
where
%
\begin{eqnarray*}
f(\omega^2)&=&\frac{3\varepsilon_{\rm i}\omega}{(\varepsilon_{\rm
r}+2)^2+\varepsilon_{\rm i}^2}>0, \qquad g(\omega^2)=
\frac{(\varepsilon_{\rm r}+2)(\varepsilon_{\rm
r}-1)+\varepsilon_{\rm i}^2}{(\varepsilon_{\rm
r}+2)^2+\varepsilon_{\rm i}^2}.
\end{eqnarray*}
The change of variables $\omega_1,\omega_2\mapsto\omega_\bot, \phi$:
%
\begin{equation}
\omega_1=\omega_\bot\cos\phi,\qquad \omega_2=\omega_\bot\sin\phi
\lab{3.11}
\end{equation}
reduces the set \re{3.10} to quadratures:
%
\begin{eqnarray}
\tau&=&-\int\frac{\D{}{\omega_\bot^2}\,\omega^2}{2\omega_\bot^2\,f(\omega^2)},
\lab{3.12}\\
\phi&=&-\omega_3\int\D{}{\tau}\frac{g(0)-g(\omega^2)}{\omega^2}\nn\\
&=&\omega_3\int\frac{\D{}{\omega_\bot^2}\,(g(0)-g(\omega^2))}{2\omega_\bot^2\,f(\omega^2)}.
\lab{3.13}
\end{eqnarray}

The vector $\Pc$ determining a particle polarization is inclined to
the external field vector $\Ec$ by the angle $\psi$ depending on the
components of the angular velocity $\B\omega$:
%
\begin{subequations}\lab{3.14}
\begin{eqnarray}
\tan^2\psi&=&\frac{\{f^2(\omega^2)/\omega^2+[g(0)-g(\omega^2)]^2\omega_3^2\}\omega_\bot^2}{\{g(\omega^2)+[g(0)-g(\omega^2)]\omega_3\}^2},
\lab{3.14a}\\
\tan\psi&\le&\tan\psi|_{\omega_3=0}=f(\omega^2)/g(\omega^2)
\lab{3.14b}
\end{eqnarray}
\end{subequations}
since given $\omega^2$, the maximal value of $\psi$ corresponds to
$\omega_3=0$.

\subsection{Debye relaxation model of polar dielectric}

The simplest nontrivial form of the dielectric function arises from
the Debye relaxation theory of polar dielectrics \cite{Deb29}. The
material equation \re{3.1} in this case reads:
%
\begin{equation}
\B D(t)=\B E(t)+\int\limits_{-\infty}^{t}\D{}{t'}\chi(t-t')\B
E(t'),\lab{3.15}
\end{equation}
where $\chi(t)=\chi_0\,\mathrm{e}^{-t/\tau_0}$ and $\tau_0$ is the
relaxation time. With a commonly used minimal generalization
\cite{Deb29,Raj16} the corresponding dielectric function has the
form:
%
\begin{equation}
\varepsilon(\Omega)=\varepsilon_\infty + \frac{\varepsilon_{\rm
s}-\varepsilon_\infty}{1-\mathrm{i}\tau_0\Omega}, \lab{3.16}
\end{equation}
where $\varepsilon_{\rm s}=\varepsilon(0)$ is the static dielectric
permittivity, and higher frequency absorption mechanisms are
accounted by the phenomenological constant $\varepsilon_\infty$; in
general, $\varepsilon_\infty\ne1$.

In terms of \re{2.27}, \re{2.28}, \re{3.5} we have:
%
\begin{eqnarray*}
\frac{\varepsilon_{\rm i}}{\omega}=\frac{\varepsilon_{\rm
s}-\varepsilon_\infty}{1+\omega^2},\qquad \varepsilon_{\rm
r}=\varepsilon_\infty + \frac{\varepsilon_{\rm i}}{\omega}.
\end{eqnarray*}
Inserting these functions into the quadratures \re{3.12}, \re{3.13},
one can derive the latter in the form:
%
\begin{eqnarray}
-6(\varepsilon_{\rm s}-\varepsilon_\infty)\tau&=&\{(\varepsilon_{\rm
s}+2)^2+(\varepsilon_\infty+2)\omega_3^2\}\ln\omega_\bot^2 +
(\varepsilon_\infty+2)\omega_\bot^2;
\lab{3.17}\\
\phi&=&\omega_3\frac{\varepsilon_\infty+2}{\varepsilon_{\rm
s}+2}\ln\omega_\bot^2. \lab{3.18}
\end{eqnarray}
The change of the variable
%
\begin{eqnarray*}
\omega_\bot^2\mapsto\nu&=&\frac{(\varepsilon_\infty+2)^2}{3(\varepsilon_{\rm
s}-\varepsilon_\infty)\xi}\,\omega_\bot^2,\qquad
\mbox{where}\quad\xi=\frac{(\varepsilon_{\rm
s}+2)^2+(\varepsilon_\infty+2)\omega_3^2}{3(\varepsilon_{\rm
s}-\varepsilon_\infty)}
\end{eqnarray*}
reduces the solution \re{3.17} to the form \re{2.32}, and then to
the explicit one \re{2.33}. Asymptotically, at $t\to\infty$, we
have:
%
\begin{equation}
\Omega_\bot\sim\mathrm{e}^{-t/T},\qquad
\phi\sim-\frac{(\varepsilon_\infty+2)\tau_0\Omega_3
t}{(\varepsilon_{\rm s}+2)T}, \lab{3.19}
\end{equation}
where
%
\begin{equation}
T(\Omega_3)=\frac{8\pi}{45}\,\frac{(\varepsilon_{\rm
s}+2)^2+(\varepsilon_\infty+2)^2\tau_0^2\Omega_3^2}{\varepsilon_{\rm
s}-\varepsilon_\infty}\, \frac{\mu R^2}{\tau_0\mathcal{E}^2}
\lab{3.20}
\end{equation}

\subsubsection{Numerical example: water ice}
Going over to a numerical example, it is worth to note that the
relaxation model \re{3.15} is appropriate for the description of
polar liquids. Nevertheless, Debye shown \cite{Deb29} that the
dielectric function \re{3.16} describes satisfactory a water ice,
the polar solid. We consider here this example, using data from
\cite{Deb29}, corresponding to the ice temperature -2$^\circ$C:
$\mu=0.9\,$g/cm$^3$, $\varepsilon_{\rm s}=80$,
$\varepsilon_\infty=2.2$, $\tau_0=4.3\cdot10^{-5}\,$s.\footnote{
This value of $\tau_0$ is the indicated in \cite{Deb29} value of the
settled life time of a molecule $\tau_{\rm l}=2.2\cdot10^{-6}\,$s
multiplied by the factor $(\varepsilon_{\rm
s}+2)/(\varepsilon_\infty+2)$.} Then for the record initial angular
velocity $\Omega_{\rm max}=2\pi\cdot10^9\,$s$^{-1}$ its
dimensionless value $\omega\le\omega_{\rm max}=\tau_0\Omega_{\rm
max}=2.7\cdot10^5$. Thus rotary effects are expected essential.
Moreover, they depend strongly on the initial direction of $\BOm$
with respect to $\Ec$, i.e., on the value of
$\omega_3=\tau_0\Omega_3$. In the case $\omega_3\approx\omega_{\rm
max}=2.7\cdot10^5$ the braking time $T=4.34\cdot10^{-2}\,$s, thus
$\tau_0\ll T$ and the adiabatic approximation is appropriate. If
$\omega_3\lesssim1$, the braking time decreases up to
$T=2.3\cdot10^{-8}\,$s, and the validity of the adiabatic
approximation becomes questionable. The initial inclination angle
may approach $\psi_{\rm max}\lesssim\pi/2$, i.e., nearly
$\Pc\bot\Ec$.

The relaxation time of the {\em ice} appearing in the Debye
dielectric function is unusually long compared to that of, say, {\em
liquid water}, $\tau_0=5\cdot10^{-10}\,$s  at 20$^\circ$C
\cite{Deb29}. If the water droplet rotates rigidly with the angular
velocity $\Omega\le\Omega_{\rm max}=2\pi\cdot10^9\,$s$^{-1}$, the
inclination angle $\psi\lesssim20^\circ$ is rather perceivable; it
decreases in the braking time $T\approx T_0\approx2\cdot10^{-3}\,$s.
But these figures are very approximate, since the deformation of the
droplet by centrifugal and electrostatic forces is not taken into
account.

The same or less order of the relaxation time
$\tau_0\sim10^{-11}-10^{-12}\,$s is characteristic for the polyvinyl
chloride and some other polymer polar solids. But the Debye
dielectric function \re{3.16} corresponding to the single relaxation
time model is not appropriate for these cases, and phenomenological
generalizations such as the Cole-Cole function or the Devidson-Cole
function should be applied \cite{Raj16}. These dielectric functions
are more complicated in calculations and are not considered in the
present paper.

\subsection{Lorentz oscillator model of non-polar dielectric}

Following the Lorentz model \cite{Lor09} electrons in dielectrics
are considered as the charged damped harmonic oscillators driven by
the electric field penetrating in the dielectric. We consider the
simplest single-resonant model based on the following dielectric
function:
%
\begin{equation}
\varepsilon(\Omega)=\varepsilon_\infty +
\frac{\Omega_p^2}{\Omega_0^2-\Omega^2 -\mathrm{i}\Gamma\Omega},
\lab{3.21}
\end{equation}
where $\Omega_p$ is the plasma frequency, $\Omega_0$ is the
resonance frequency, $\Gamma$ is the damping decrement, and the
value $\varepsilon_\infty$ of the dielectric function at infinite
frequency  (instead of 1 in true Lorentz model) is used as an
adjustable parameter for accounting higher frequency resonances.

Using the dimensionless variable and constants:
%
\begin{equation}
\omega=\Omega/\Omega_0,\qquad \omega_p=\Omega_p/\Omega_0,\qquad
\gamma=\Gamma/\Omega_0, \lab{3.22}
\end{equation}
we arrive at the relations for dielectric constants and functions:
%
\begin{eqnarray*}
\varepsilon_{\rm r}&=&\varepsilon_\infty + \frac{(\varepsilon_{\rm
s}-\varepsilon_\infty)(1-\omega^2)}{(1-\omega^2)^2+\gamma^2\omega^2},\qquad
\frac{\varepsilon_{\rm i}}{\omega}=\frac{\gamma^2(\varepsilon_{\rm
s}-\varepsilon_\infty)}{(1-\omega^2)^2+\gamma^2\omega^2},\qquad
\varepsilon_{\rm s}=\varepsilon_\infty+\omega_p^2.
\end{eqnarray*}
Inserting these into the quadratures \re{3.12}, \re{3.13}, one
derives those in the form:
%
{\small
\begin{eqnarray}
-6\gamma(\varepsilon_{\rm
s}-\varepsilon_\infty)\tau&=&\{[\varepsilon_{\rm
s}-\varepsilon_\infty+(\varepsilon_\infty+2)(1-\omega_3^2)]^2
+\gamma^2(\varepsilon_\infty+2)^2\omega_3^2\}\ln\omega_\bot^2 \nn\\
&&{} - (\varepsilon_\infty+2)\{2[\varepsilon_{\rm
s}-\varepsilon_\infty  + (\varepsilon_\infty+2)(1-\omega_3^2)] -
\gamma^2(\varepsilon_\infty+2)\}\omega_\bot^2 \nn\\
&&{} +(\varepsilon_\infty+2)^2\omega_\bot^4/2;
\lab{3.23}\\
-2\gamma(\varepsilon_{\rm s}+2)\phi&=&\omega_3\{[\varepsilon_{\rm
s}-\varepsilon_\infty+(\varepsilon_\infty+2)(1-\gamma^2-\omega_3^2)]\ln\omega_\bot^2
-(\varepsilon_\infty+2)\omega_\bot^2\}. \lab{3.24}
\end{eqnarray}
}

In contrast to pervious examples, this implicit solution cannot be
unrevealed explicitly. Actually, one can explicit the asymptotical
solution at $\tau\to\infty$:
%
{\small
\begin{eqnarray*}
\omega_\bot&\sim&\exp\left\{-\frac{3\gamma(\varepsilon_{\rm
s}-\varepsilon_\infty)\tau}{[\varepsilon_{\rm
s}-\varepsilon_\infty+(\varepsilon_\infty+2)(1-\omega_3^2)]^2
+\gamma^2(\varepsilon_\infty+2)^2\omega_3^2}\right\},
\\
\phi&\sim&3\frac{\omega_3(\varepsilon_{\rm
s}-\varepsilon_\infty)[\varepsilon_{\rm
s}-\varepsilon_\infty+(\varepsilon_\infty+2)(1-\gamma^2-\omega_3^2)]\tau}{
[\varepsilon_{\rm
s}-\varepsilon_\infty+(\varepsilon_\infty+2)(1-\omega_3^2)]^2
+\gamma^2(\varepsilon_\infty+2)^2\omega_3^2}.
\end{eqnarray*}
}
Even so, it is rather cumbersome. However, a practice corresponds
mainly to the infrared domain $\Omega/\Omega_0\ll1$ where the
$t\to\infty$ asymptotics simplifies considerably:
%
\begin{eqnarray}
&&\Omega_\bot\sim\mathrm{e}^{-t/T}, \qquad \phi\sim\tilde\Omega_3t,
\lab{3.25}\\
\mbox{where}\quad&&T=\frac{T_0(\varepsilon_\infty+2)^2}{3(\varepsilon_{\rm
s}-\varepsilon_\infty)},\quad T_0=\frac{I}{\tau_0R^3{\cal
E}^2},\quad\tau_0=\frac\Gamma{\Omega_0^2},\quad
\tilde\Omega_3=\frac{\Omega_3}{T\Gamma}. \lab{3.26}
\end{eqnarray}
In contrast to previous examples, here the precession is
counterclockwise.

\subsubsection{Numerical example: hexagonal silicon carbide}
 Following \cite{SKW59}, dielectric properties of this material can be approximated satisfactory
by the single resonant Lorentz function \re{3.21} with the
parameters: $\Omega_0=2\pi\cdot23.8\,$THz
$\,=2\pi\cdot2.38\cdot10^{13}\,$s$^{-1}$,
$\omega_p^2=\Omega_p^2/\Omega_0^2=3.305$,
$\gamma=\Gamma/\Omega_0=0.006$, $\varepsilon_\infty=6.7$, so that
$\varepsilon_{\rm s}=10.005$. Thus for even record frequency
$\Omega_{\rm max}=2\pi\cdot10^{9}\,$s$^{-1}$ the dimensionless value
$\omega\le\omega_{\rm max}=\Omega_{\rm
max}/\Omega_0=2.64\cdot10^{-4}\ll1$.

Taking into account the mass density $\mu=3.23\,$g/cm$^3$ and
familiar for this work the external field
${\mathcal{E}}=33.4\,$statV/cm one obtains
$\tau_0=2.5\cdot10^{-16}\,$s,
$T=7\cdot10^{3}\,$s$\,\approx2\,$hours, so that $\tau_0\ll T$, and
the adiabatic approximation is perfect. The time-scale quantity
$\tau_0$ is longer than that of conductors (even high resistance
ones), but much shorter then the relaxation time of the ice. The
braking time $T$ and the inclination angle $\psi\le4.8\cdot10^{-7}$
are close to those of nichrome.


\section{Poor conductors and conductor-coated dielectrics}
\renewcommand{\theequation}{4.\arabic{equation}}
\setcounter{equation}{0}

Results of previous subsections permit us to consider poor-conductor
and non-homogeneous particles. We consider a very simple model of a
poor-conductor particle or a dielectric core layered by the
well-conducting shell. The high- and low-frequency properties of the
core is characterized by the dielectric constant
$\varepsilon\equiv\varepsilon_\infty$ and the bulk conductivity
$\varkappa$, respectively, while the shell of the thickness $h\ll R$
(if any) possesses the surface conductivity $\kappa$. Dispersion of
these quantities can be but here is not taken into account.

The ansatz for the potential \re{2.17} remains valid while the
boundary condition \re{2.18} or \re{3.6} changes:
%
\begin{equation}
\sigma=\frac1{4\pi}(E_r^+-D_r^-)=\frac3{4\pi}\tilde{\cal P}_r,
\lab{4.1}
\end{equation}
where $\B D^-=\varepsilon\B E^-$  and
%
\begin{equation}
\tilde\Pc=[(\varepsilon+2)\Pc-(\varepsilon-1)\Ec]/3. \lab{4.2}
\end{equation}
Subsequently, the components ${\cal P}_y$ and ${\cal P}_x$ in
r.-h.s. of the equations \re{2.22} must be replaced formally by
$\tilde{\cal P}_y$ and $\tilde{\cal P}_x$ which yields the solution:
%
\begin{equation}
{\cal
P}_x=\frac{1+\beta\tilde\tau_0^2\Omega^2}{1+\tilde\tau_0^2\Omega^2}\,\mathcal{E}_x,
\qquad {\cal
P}_y=\frac{(1-\beta)\tilde\tau_0\Omega}{1+\tilde\tau_0^2\Omega^2}\,\mathcal{E}_x,
\qquad {\cal P}_z=\mathcal{E}_z, \lab{4.3}
\end{equation}
with
%
\begin{equation}
\tilde\tau_0
=\frac{\tau_0}{1-\beta}=\frac{\varepsilon+2}{4\pi(\varkappa+2\kappa/R)},
\quad \beta=\frac{\varepsilon-1}{\varepsilon+2}, \lab{4.4}
\end{equation}
instead of \re{2.21}, \re{2.22}. Then the torque in r.-h.s. of the
Euler equation \re{2.25} or \re{2.26} must be replaced by the
expression:
%
\begin{equation}
\B
M=\frac{(1-\beta)\tilde\tau_0R^3}{1+\tilde\tau_0^2\Omega^2}\Ec\times\{\Ec\times\BOm-\tilde\tau_0(\BOm\cdot\Ec)\BOm\}.
\lab{4.5}
\end{equation}
The transition to the dimensionless variables
%
\begin{equation*}
\B\omega=\tilde\tau_0\BOm, \qquad \tau=t/T_0,
\end{equation*}
where
%
\begin{equation*}
T_0=\frac{I}{(1{-}\beta)\tilde\tau_0R^3\mathcal{E}^2}=\frac{I}{\tau_0R^3\mathcal{E}^2},
\end{equation*}
reduces formally the Euler equation \re{2.25} to the same
dimensionless component set \re{2.29}--\re{2.31}. Let us note that
the braking time is $T=T_0(1+\omega_3^2)$ where $T_0$ does not
depend on the dielectric permittivity of the core, but only on its
and surface conductivity. The inclination angle is determined by the
equation:
%
\begin{equation*}
\tan\psi=\frac{(1-\beta)\sqrt{1+\omega_3^2}\,\omega_\bot}{|1+\beta\omega^2+(1-\beta)\omega_3|}
\le\tan\psi|_{\omega_3=0}=\frac{(1-\beta)\omega}{1+\beta\omega^2}.
\end{equation*}

\subsection{Numerical examples}

\subsubsection{LISICON particle}
Entrapment and study of levitating particles of solid electrolytes
may appear useful for improving characteristics of these materials.
Here we consider $R=50\,$nm particles of the commonly used LISICON
solid electrolyte \cite{MBM84,Zh..A23}. Taking into account the mass
density $\mu=2.8\,$g/cm$^3$, dielectric permittivity
$\varepsilon_\infty=21.4$ and the conductivity
$\varkappa=4.52\cdot10^{12}\,$s$^{-1}$ (i.e., 502 S/m in SI)
achieved in certain samples at temperature 250$^\circ\,$C, one
obtains the relaxation time $\tilde\tau_0=4.1\cdot10^{-10}\,$s.
Since $\omega\le\tilde\tau_0\Omega_{\rm max}=2.6$, rotary effects
are not negligible.  In particular, the inclination angle
$\psi\lesssim2.8^\circ$. The braking time $T_0=0.002\,$s$\ \le
T_{\Omega_3\approx\Omega_{\rm max}}=0.015\,$s, so the adiabatic
approximation is reliable.

\subsubsection{Gold-coated silica particle}
Synthesis of silica nanoparticles covered by gold shell is currently
elaborated \cite{F..S21,T..A23} for physical, biological and medical
use. We consider such a particle of the size $R=50\,$nm (closed to
those of \cite{F..S21}) with the fused silica core and the $h=3\,$nm
gold shell. Taking into account the mass density
$\mu=2.2\,$g/cm$^3$, dielectric permittivity $\varepsilon=3.8$ and a
practically zero conductivity of the fused silica, and corresponding
data from Subsection 2.4 for the gold, one obtains the relaxation
time $\tilde\tau_0=10^{-17}\,$s. Since
$\omega\le\tilde\tau_0\Omega_{\rm max}=6.45\cdot10^{-8}$, the
braking time is $T=T_0(1+\tilde\tau_0^2\Omega_3^2)\approx
T_0=6.8\cdot10^5\,$s $\approx13.3\,$hours $\gg\tilde\tau_0$. Thus
the adiabatic approximation is excellent while the inclination angle
$\psi\lesssim6.45\cdot10^{-8}$  is tiny, similarly to the case of
entirely golden particle.


\section{Conclusion and application perspective}
\renewcommand{\theequation}{5.\arabic{equation}}
\setcounter{equation}{0}

Free rotation of neutral spherical particles levitating in the
uniform electrostatic field $\Ec$ has been considered. The external
field $\Ec$ induces in particles the electric dipole moment $\B{d}$
which is inclined to this field due to particle rotation $\BOm$. In
turn, the interaction of the particle dipole moment $\B{d}$ with the
external field $\Ec$ causes the torque $\B M\equiv\B{d}\times\Ec$
braking the particle rotation.

Three basic examples has been considered: the Ohm conductor, the
Debye model of polar dielectric, and the Lorentz model of non-polar
dielectric. Besides, the hybrid conductor-dielectric is also
included. We assume that the distribution of free or bound charges
resulting in a dipole moment follow adiabatically the change of
particle rotation with the characteristic relaxation time $\tau_0$.
The corresponding Euler equations of particle rotary motion are
reduced to quadratures and integrated out. In all cases solutions
reveal common features.

The parallel to the external field $\Ec$ component of angular
velocity is unchanged,  $\Omega_\|=\,$const, while the orthogonal
component $\Omega_\bot$ decreases asymptotically (at $t\to\infty$)
by the exponential law $\Omega_\bot\sim\exp\{-t/T\}$ with the
characteristic braking time:
%
\begin{equation}\label{5.1}
T(\Omega_\|)\ge T(0)\equiv
T_0\sim\frac{I}{\tau_0R^3\mathcal{E}^2}\sim\frac{\mu
R^2}{\tau_0\mathcal{E}^2}.
\end{equation}
The same happens with the dipole moment:  ${d}_\|=\,$const while
$d_\bot\sim\exp\{-t/T\}$.

This behavior differs strongly from that of the Quincke effect in
despite the electromechanical description of both cases is common.
The difference is that the particle in our case is surrounded by
vacuum while in Quincke's case it is immersed in a conducting
viscous medium in which $\Omega_\|=0$ while $\Omega_\bot$ and
$d_\bot$ are regarded as constants \cite{M-T69,Jon84}. Actually, the
initial magnitude and direction of $\BOm$ can be arbitrary in both
cases. But the viscous conducting medium dumps $\Omega_\|$ quickly
to zero and provides the particle polarization and corresponding
torque opposite to that in our case. Consequently, $\Omega_\bot$
evolves to non-zero asymptotical value at which the viscous torque
compensates the electric one. This stationary state reached after
transient processes is a stage of the Quincke effect, and its 2D
description is sufficient.

In our case, namely the transition process of a particle 3D rotation
is of interest since the estimated braking time in vacuum \re{5.1}
may occur notable. At least, $T\gg\tau_0$ for the adiabatic
approximation to be valid. Details of this process depend on
electric properties of particles.

The value of the relaxation time $\tau_0$ varies in wide range, from
$\tau_0\sim6\cdot10^{-19}\,$s for good conducting golden particles
to $\tau_0\sim5\cdot10^{-5}\,$s for ice, the polar dielectric. The
same is true for the maximal inclination angle at $\Omega_{\rm
max}=2\pi\,$GHz: from $\psi_{\rm max}\sim4\cdot10^{-9}$ to
$\psi_{\rm max}\sim\pi/2$. Ceteris paribus, the minimal braking time
$T_0$ is inversely proportional to the relaxation time $\tau_0$,
thus it varies in wide range too: from $T_0\sim10^{6}\,$s for golden
particles to $T_0\sim2\cdot10^{-8}\,$s for ice. Despite such a large
difference in numbers, the solutions for Ohm conductors and Debye
polar dielectrics are similar: up to a replacement of parameters,
they coincide and can be represented in an explicit form via the
Lambert W-function. In particular, they describe the clockwise
precession with the angular velocity
$\tilde\Omega_\|\sim\Omega_\|\tau_0/T$ of the vector $\BOm$ when
turning to its asymptotical value $\Omega_\|\Ec/{\cal E}$.

In contrast, the simplest Lorentz model of non-polar dielectric
characterized by the single resonance frequency $\Omega_0$ leads to
more complicated solution in implicit parametric form which
describes the counterclockwise precession with the angular velocity
$\tilde\Omega_\|\sim\Omega_\|(T\Gamma)^{-1}$ in the infrared region
$\Omega<\Omega_0$. The reason is that the real part ${\rm
Re}\,\varepsilon(\Omega)$ of the Lorentz dielectric function
\re{3.21} is increasing in this domain, in contrast to the Debye
function \re{3.16}.

Presented analytical solutions of the equations of rotary motion may
have an application in the modern particle trap physics.

The motion of charged particles in the ideal Penning trap is
described by an exact analytical solution of the corresponding
equations of motion. This gives one possible to account imperfection
effects as perturbations \cite{Vog18,YPM15}.

The capture of neutral particles is carried out using the
interaction of their dipole moment with the electromagnetic field.
For example, in the currently designed trap for neutral polar
particles \cite{PMY2020} the static sixtupolar electric and
quadrupolar magnetic confining fields are complemented by a strong
uniform electric field intended to orient permanent dipole moments
of particles along a symmetry axis of the device for better
trapping. Equations of particle motion in this trap tangle
translational and rotational degrees of freedom and are, in general,
non-integrable.

The strong orientational electric field can be used not only for
manipulating polar particles but also for inducing dipole moment in
non-polar particles. The authors of the new design trap
\cite{PMY2020} consider the orientational electric field of order
0.1--1 V/m. But its strength is not restricted from above
principally. Thus, a much stronger field can be assumed (in
principle, up to the order of $10^6\,$V/m). Then namely this field
in the first approximation will determine a rotary motion of
particles while the confining fields can be accounted as
perturbations. The problem raised in the present paper can be used
to split approximately rotational degrees of freedom from
translational ones and thus to simplify the analysis of the new
design trap.

As it is noted above, the external electrostatic field suppresses,
by dissipation, those components of the particle angular velocity
which are transverse to the field. Therefore, the effect of the
braking is not only the alignment of the induced dipole moment along
the field, but also of the angular momentum of the particle,
regardless of its initial rotation. For this, the particles do not
necessarily have to levitate: the alignment can occur during the
free fall of neutral particles in the field, if the braking time is
small enough (as for some polar dielectrics or solid electrolytes).
Therefore, the described effect may extend the perspectives of the
aforementioned optomechanical experiments
\cite{RDHD18,AXBJGL20,JYRLYZ21,S-B16}.

\small

\end{document}